# Near-infrared photoluminescence from molecular crystals containing tellurium

**Hong-Tao Sun,*[ab] Yoshio Sakka,[c] Naoto Shirahata,[c,d] Minoru Fujii,[e] and Tetsu Yonezawa[a]**

5  We report the observation of near-infrared photoluminescence from $Te_4(Ga_2Cl_7)_2$ and $Te_4(Al_2Cl_7)_2$ molecular crystals containing $Te_4^{2+}$ polycations. The experimental and theoretical results clearly revealed that $Te_4^{2+}$ polycation is one smart near-infrared emitter with characteristic emission peaks at 1252 and 1258 nm for $Te_4(Ga_2Cl_7)_2$ and $Te_4(Al_2Cl_7)_2$ crystals, respectively, resulting from the intrinsic electronic transitions of $Te_4^{2+}$. Furthermore, it was also found that the emissions strongly depend on the excitation
10  wavelengths for both $Te_4(Ga_2Cl_7)_2$ and $Te_4(Al_2Cl_7)_2$ samples, most possibly owing to the co-existence of other Te-related optically active centers. This research not only enriches the species of luminescent charged p-block element polyhedra and deepens the understanding of Te-related photophysical behaviors, but also may stimulate efforts for designing novel material systems using such active centers. It is also greatly expected that these sub-nanometer optically active species could exist in other systems such
15  as glasses, polymers, and bulk optical crystals, and the stabilization of these centers in widely used hosts will pave the way for their practical applications.

## Introduction

The development of inorganic luminescent materials (ILMs) that lack the intrinsic limitations of organic fluorophores is an area of considerable current interest across a number of science, engineering and biomedical disciplines.[1] The luminescence from these materials results from the electronic transitions of active centers or defects in inorganic matrix, quantum-size confinement as in the case of nanocrystals or exciton recombination of semiconductors. To date, ILMs have witnessed an explosion of applications in a broad variety of areas such as telecommunication, lighting, displays, information technology, and nano-biomedicines. For instance, light-emitting semiconductor p-n junctions, carefully engineered structures that bring electrons and holes together to generate light, has been a vital and indispensable building block of modern optoelectronics, particularly for creating laser diodes and LEDs. In view of the impressive achievements and breakthroughs obtained by using these traditional ILMs, it is believed that developing new type of photonic material systems would not only enrich the well established spectrum of ILMs, but might open up new possibilities for their functional applications and for exploring the unique photophysical characteristics inherent to them.

As is well known, heavier p-block elements have fundamentally different electronic properties from the lighter congeners. It has been revealed that these elements could form fascinating structures such as anionic and cationic polyhedrons, which are considered to be remarkable species domiciled in solid-state chemistry.[2-10] One of the most important aspects of these structures is that they provide a window into the rather ill-defined area of chemistry that lies between isolated molecular species and solid-state compounds with extended structures.[4e] Over the past decades, particular interests have been focused on Group 13, 14 and 15 charged polyhedra and great progress on the synthesis, experimental characterization and theoretical investigations of related compounds such as NaSi, NaGeK$_4$, Rb$_4$Sn$_4$(NH$_3$)$_2$, Rb$_4$Pb$_4$(NH$_3$)$_2$, Bi$_5$(AlCl$_4$)$_3$, Bi$_5$(GaCl$_4$)$_3$, Bi$_8$(Al$_2$Cl$_7$)$_2$, and [Bi$_{10}$Au$_2$](SbBi$_3$Br$_9$)$_2$ has been made.[2-8] However, the preparation as well as systematical evaluation of physicochemical properties of compounds containing the clusters of Group 16 elements has not attracted deserved attention in the research community.[9]

Herein, we demonstrate the first photophysical study of luminescent Te$_4$(Ga$_2$Cl$_7$)$_2$ and Te$_4$(Al$_2$Cl$_7$)$_2$ molecular crystals containing Te$_4^{2+}$ polycations. The products were characterized by extensive approaches including powder X-ray diffraction (PXRD), diffuse reflectance, excitation-emission matrix (EEM) and time-resolved photoluminescence (PL) spectroscopy. Furthermore, we employed time-dependent density functional theory (TDDFT) to determine energies and compositions of excited states of Te$_4^{2+}$. The experimental and theoretical results clearly revealed that Te$_4^{2+}$ polycation is one smart near-infrared emitter with characteristic emission peaks at 1252 and 1258 nm for Te$_4$(Ga$_2$Cl$_7$)$_2$ and Te$_4$(Al$_2$Cl$_7$)$_2$ crystals, respectively. We also found that the emissions strongly depend on the excitation wavelengths, most possibly resulting from the co-existence of other Te-related optically active centers.

## Experimental details

Tellurium (Wako, 99.99%), TeCl$_4$ (Sigma-Aldrich, 99%), AlCl$_3$ (Sigma-Aldrich, 99.99%) and anhydrous GaCl$_3$ (Sigma-Aldrich, 99.999%) were used as received. To synthesize Te$_4$(Ga$_2$Cl$_7$)$_2$ crystal, Te, TeCl$_4$, and GaCl$_3$ were mixed in a 3:1:4 molar ratio in a glove box (<2 ppm H$_2$O; <0.1 ppm O$_2$), degassed and sealed in an ampule. The mixture was first kept at 250 °C for 48 hours and then slowly cooled down to 80 °C with a rate of 1°C/h. Te$_4$(Al$_2$Cl$_7$)$_2$ crystal was synthesized according to a method first reported by Couch et al.,[9a] and the mixture was thermally treated using the same procedure applied for Te$_4$(Ga$_2$Cl$_7$)$_2$, since the related information is not available in ref. 9a. The obtained products were transferred to other bottles or capillaries in a glove box for the following measurements. The products were characterized by X-ray diffractometer (Rigaku-RINT Ultima3, λ=1.54056 Å). The samples were sealed in 1 mm Hilgenberg borosilicate capillaries and kept spinning during the measurement. Owing to the high moisture sensitivity of the products, we measured the absorption spectrum of the crystals enclosed between two 1 mm thick pure silica pieces using a UV-vis-NIR spectroscope (V-570, JASCO, Japan) equipped with an integrating sphere. Photoluminescence (PL) spectra was taken by a Horiba NanoLog spectrofluorometer equipped with a monochromated Xe lamp and a liquid N$_2$ cooled photomultiplier tube (PMT) (Hamamatsu, R5509-72). Emission spectra were taken in 2 nm steps at different excitation wavelengths from 300 to 880 nm with 4 nm intervals. It is noteworthy that all spectra were corrected for spectral response of the detection system. Time-resolved PL measurements were performed by detecting the modulated luminescence signal with a PMT (Hamamatsu, R5509-72), and then analyzing the signal with a photon-counting multichannel scaler. The excitation source for the time-resolved PL measurements was 480 nm light (pulse width: 5nsec; frequency: 20Hz) from an optical parametric oscillator pumped by the third harmonic of a Nd:YAG laser. The detailed quantum chemistry calculation on the UV-Vis-NIR absorption spectra of $Te_4^{2+}$ polycation will be discussed below.

### Results and discussions

To evaluate the crystallinity of the obtained products, first we took PXRD measurement. As displayed in Fig. 1, the diffraction pattern of $Te_4(Ga_2Cl_7)_2$ corresponds well to the simulated diffractogram determined from a $Te_4(Ga_2Cl_7)_2$ single crystal isolated from $GaCl_3$-benzene media.[10] $Te_4(Al_2Cl_7)_2$ is isotypic with $Te_4(Ga_2Cl_7)_2$ (Fig. 2), both of which consist of $Te_4^{2+}$ polycation and two corner-sharing $ACl_4$ (A is Al or Ga) tetrahedra. The formation of $Te_4(Al_2Cl_7)_2$ crystal was also confirmed by PXRD (Fig. 1). However, it is worth to note that five peaks located at 11.96, 14.92, 16.29, 18.13, and 20.41° can not be ascribed to $Te_4(Al_2Cl_7)_2$ phase. After thorough examination of the diffraction patterns of related compounds, it is believed that the peak at 11.96° results from the (200) reflection of the $TeCl_4$ phase, while the peaks at 14.92, 16.29, 18.13, and 20.41° are attributable to (002), (200), (112), and (220) reflections of $Te_4(AlCl_4)_2$ phase (CSD no: 421883), respectively. Compared with the result in ref. 9a, different thermal treatment process adopted here may be one main reason for the appearance of the impurities in $Te_4(Al_2Cl_7)_2$ sample.

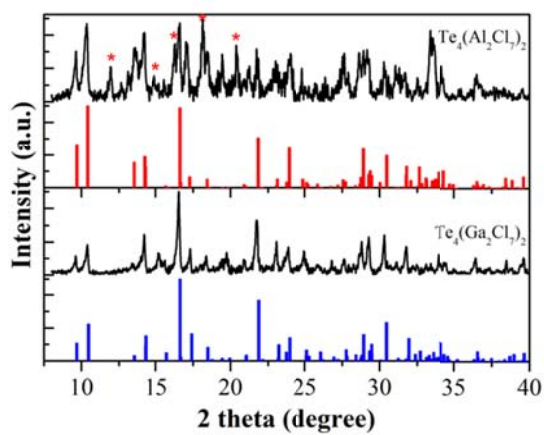

**Fig. 1** PXRD patterns of $Te_4(Ga_2Cl_7)_2$ and $Te_4(Al_2Cl_7)_2$ samples. The vertical blue and red lines denote the intensity and position of the diffraction peaks for $Te_4(Ga_2Cl_7)_2$ and $Te_4(Al_2Cl_7)_2$ phases, respectively. The peaks denoted by red asterisks are ascribed to the reflections of the impurities.

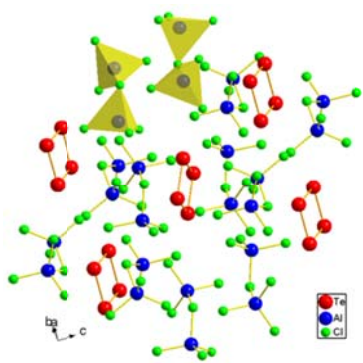

**Fig. 2** Molecular structure of $Te_4(Al_2Cl_7)_2$ in ball-stick and polyhedral representations. This crystal consists of $Te_4^{2+}$ and $[Al_2Cl_7]^-$ units, which occurs in the monoclinic space group $P2_1/c$ with $a$ = 9.113 (4), $b$ = 11.151 (6), c = 13.096 (5) Å, and β = 90.20 (2)°.

To obtain the absorption characteristics in the visible and NIR spectral ranges, we took the diffuse reflectance spectra of these samples. As shown in Fig. 3a, both samples demonstrate similar absorption bands, two visible bands peaking at 424 and 525 nm and a broad NIR band at 885 nm. In addition, it was observed that the absorption of these samples increase dramatically at < 360 nm (Fig. 3). The similarity in the spectral shapes and peak positions suggests that the electronic transitions in these samples are mainly dominated by $Te_4^{2+}$ polycations, although there remains some difference for $Te_4^{2+}$ units.[9,10] To assign the oberved absorption bands to specific electronic transitions of $Te_4^{2+}$ polycations, next we carried out quantum chemistry calculations using the Amsterdam Density Functional (ADF) program package developed by Baerends et al.[11] The experimentally determined geometries of $Te_4^{2+}$ in $Te_4(Ga_2Cl_7)_2$ and $Te_4(Al_2Cl_7)_2$ single crystals were used for the following calculations.[9,10] Spin-restricted time-dependent density functional theory (TDDFT) was employed to determine energies and compositions of excited states of these units, using the hybrid Becke three-parameter Lee−Yang−Parr (B3LYP) functional.[12] The Slater-type all-electron basis set utilized in the TDDFT calculations is of triple-ζ-polarized (TZP) quality. The absorption bands corresponding to 10 allowed and forbidden excited states were calculated through a Davidson method, and the absorption spectra were fitted with a Gaussian function with a width at half-maximum of 50 nm. The ADF numerical integration parameter was set to 5.0 in both calculations. Spin−orbit coupling was taken into account for all TDDFT calculations. We then thoroughly compared the computed UV-vis-NIR absorption spectra of $Te_4(Ga_2Cl_7)_2$ and $Te_4(Al_2Cl_7)_2$ with the original experimental absorption spectra. As shown in Fig. 3 and Table 1 and 2, two absorption bands in the visible and NIR spectral regions were determined theoretically for either $Te_4(Ga_2Cl_7)_2$ or $Te_4(Al_2Cl_7)_2$ sample; the NIR bands agree quite well with the experiment, especially in the peak position, although the computed/fitted curves are narrower than experimental ones. In contrast, the calculated visible bands are red-shifted relative to the experimental bands. Obviously, the theoretical absorption in the visible region is much stronger than that in the NIR region. This feature is in good agreement with the experimental absorption curves (Fig. 3). Interestingly, theoretical results clearly revealed that $Te_4^{2+}$ units in $Te_4(Ga_2Cl_7)_2$ and $Te_4(Al_2Cl_7)_2$ demonstrate similar characteristics of electronic transitions, resulting from the inherent structural similarity between them, which agrees well with the experimental observation. Additionally, TDDFT calculation allows us to obtain the electric dipole radiative lifetimes for some specific transitions. For instance, the transitions at ca. 960 nm have lifetimes of tens of milliseconds. It is worth to note that such theoretical lifetimes correspond to separate $Te_4^{2+}$ polycation without considering the concentration quenching or nonradiative cross relaxation.

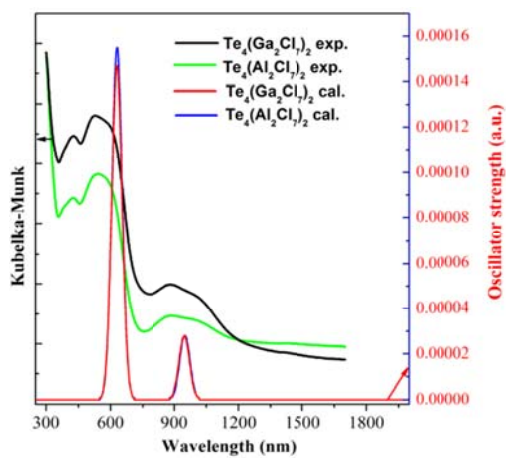

**Fig. 3** Experimentally determined diffuse reflectance spectra of $Te_4(Ga_2Cl_7)_2$ and $Te_4(Al_2Cl_7)_2$ products recorded at room temperature (black and green curves), and theoretically calculated absorption spectra of $Te_4^{2+}$ units (red and blue curves).

**Table 1.** The calculated electronic transitions of $Te_4^{2+}$ polycation in $Te_4(Ga_2Cl_7)_2$.

| No | Wavelength (nm) | Oscillator strength | Lifetime (second) | Symmetry |
|---|---|---|---|---|
| 1 | 961.8 | $6.41 \times 10^{-7}$ | $2.16 \times 10^{-2}$ | B2.u |
| 2 | 953.5 | $1.08 \times 10^{-5}$ | $1.26 \times 10^{-3}$ | B1.u |
| 3 | 953.1 | $6.62 \times 10^{-7}$ | $2.05 \times 10^{-2}$ | B3.u |
| 4 | 951.4 | 0 | | A.u |
| 5 | 944.9 | $1.63 \times 10^{-5}$ | $8.18 \times 10^{-4}$ | B1.u |
| 6 | 942.3 | 0 | | A.u |

| 7 | 656.3 | 1.31 x 10<sup>-6</sup> | 4.92 x 10<sup>-3</sup> | B3.u |
| 8 | 652.4 | 2.35 x 10<sup>-7</sup> | 2.71 x 10<sup>-2</sup> | B2.u |
| 9 | 630.5 | 1.47 x 10<sup>-4</sup> | 4.05 x 10<sup>-5</sup> | B1.u |
| 10 | 621.6 | 0 | | A.u |

**Table 2.** The calculated electronic transitions of $Te_4^{2+}$ polycation in $Te_4(Al_2Cl_7)_2$.

| No | Wavelength (nm) | Oscillator strength | Lifetime (second) | Symmetry |
|---|---|---|---|---|
| 1 | 961.3 | 6.69 x 10$^{-7}$ | 2.06 x 10$^{-2}$ | B2.u |
| 2 | 958.9 | 6.43 x 10$^{-7}$ | 2.14 x 10$^{-2}$ | B3.u |
| 3 | 952.9 | 0 | | A.u |
| 4 | 951.6 | 1.94 x 10$^{-5}$ | 6.98 x 10$^{-4}$ | B1.u |
| 5 | 949.1 | 7.71 x 10$^{-6}$ | 1.75 x 10$^{-3}$ | B1.u |
| 6 | 948.4 | 0 | | A.u |
| 7 | 656.2 | 4.45 x 10$^{-7}$ | 1.45 x 10$^{-2}$ | B2.u |
| 8 | 654.8 | 4.90 x 10$^{-7}$ | 1.31 x 10$^{-2}$ | B3.u |
| 9 | 630.9 | 1.55 x 10$^{-4}$ | 3.84 x 10$^{-5}$ | B1.u |
| 10 | 616.7 | 0 | | A.u |

Fluorescence excitation–emission matrix (EEM) spectroscopy is a powerful and widely used technique for characterizing the fluorescent organic or inorganic systems. This method entails collection of emission spectra at multiple excitation wavelengths, resulting in topographic or contour plots where the excitation and emission coordinates of observable fluorescence peaks can be used to identify different fluorophore moieties. Next, we employed this technique to thoroughly evaluate the steady-state PL properties of $Te_4(Ga_2Cl_7)_2$ and $Te_4(Al_2Cl_7)_2$ samples, as demonstrated in Fig. 4a and 4b, respectively. Both samples show similar emission features when excited in the range of 400-680 nm. Under the excitation of 470 and 620 nm, broad emissions peaking at 1252 and 1258 nm for $Te_4(Ga_2Cl_7)_2$ and $Te_4(Al_2Cl_7)_2$ samples, respectively, can be observed (Fig. 5), of which the full width at the half maximum (FWHM) is 240 nm for $Te_4(Ga_2Cl_7)_2$ and 258 nm for $Te_4(Al_2Cl_7)_2$. Further detailed examination revealed that the PL lineshapes are almost identical for both samples, suggesting that similar emitters in both samples contribute to the observed emission discussed above. Clearly, both samples contain $Te_4^{2+}$ polycations, which show absorption bands in the range of 400-680 nm as experimentally and theoretically revealed (Fig. 2 and Tables 1 and 2). This feature is consistent with the PL results since broad excitation bands occur in this range. These emission bands, therefore, should result from the inherent electronic transitions of $Te_4^{2+}$ polycations. However, the samples demonstrate rather different emission behaviors when the excitation wavelength shifts to the UV spectral range (Figs. 4 and 5). At 340 nm excitation, the emission lineshape from $Te_4(Ga_2Cl_7)_2$ sample does not change relative to those at 470 and 620 nm, which leads us to infer that one kind of emitter (i.e., $Te_4^{2+}$) for $Te_4(Ga_2Cl_7)_2$ contributes to the NIR emission in the excitation range of 300-680 nm. In contrast, a new emission band with peaks at ca. 1167 and 1086 nm appears for $Te_4(Al_2Cl_7)_2$ sample (Fig. 5b), indicating that other emitters could be excitable at this wavelengh. In addition to the peaks mentioned above, it is observed that both samples show PL signals when excited in the range of 680-880 nm, although the signal to noise ratios are rather poor (Supporting information, Figs. S1 and S2). $Te_4^{2+}$ polycations have characteristic absorption bands in this range, but it is obvious that these signals can not merely be assigned to such units, given that the fingerprint emission of $Te_4^{2+}$ appears in the range of 1000-1600 nm with a peak at 1252 or 1258 nm (Fig. 5).

At present, the exact mechanism of the PL bands except for that peaking at ca. 1260 nm remain unclear. In view of the synthesis procedure used here, it is likely that the species could be some Te-related emitters. As is known, Te can form radical species of $Te_2$ and $Te_2^-$ in some inorganic compounds such as zeolite-based materials.[13] It was reported that $Te_2$ and $Te_2^-$ existing in ultramarine-type solids display visible absorption bands at ca. 400 and 590 nm, respectively.[14] Both $Te_2$ and $Te_2^-$ are optically active, resulting in emissions peaking at 562 and 862 nm, respectively.[14] As shown in Fig. 4, $Te_4(Ga_2Cl_7)_2$ and $Te_4(Al_2Cl_7)_2$ samples demonstrate emissions beyond 1000 nm, thus leading us to conclude that $Te_2$ and $Te_2^-$ are not the contributors of the emissions observed here. Recently, luminescence from Te doped glasses have also been observed.[15,16] For instance, Punpai et al. found that Te-doped soda-lime-silicate glasses prepared in a reducing atmosphere show broad NIR emission with a peak at 1200 nm, which is similar to the emission band of $Te_4(Al_2Cl_7)_2$ sample at 340 nm excitation (Fig. 5b); clearly, it can not be assigned to $Te_2$ or $Te_2^-$ emitter.[13,14] Very recently, Dianov et al. reported that Te doped glass fiber shows broad NIR emission peaking at 1500 nm.[16] These experimental facts suggest that, in addition to $Te_4^{2+}$ polycation identified here, there are some other kinds of Te-related active centers.

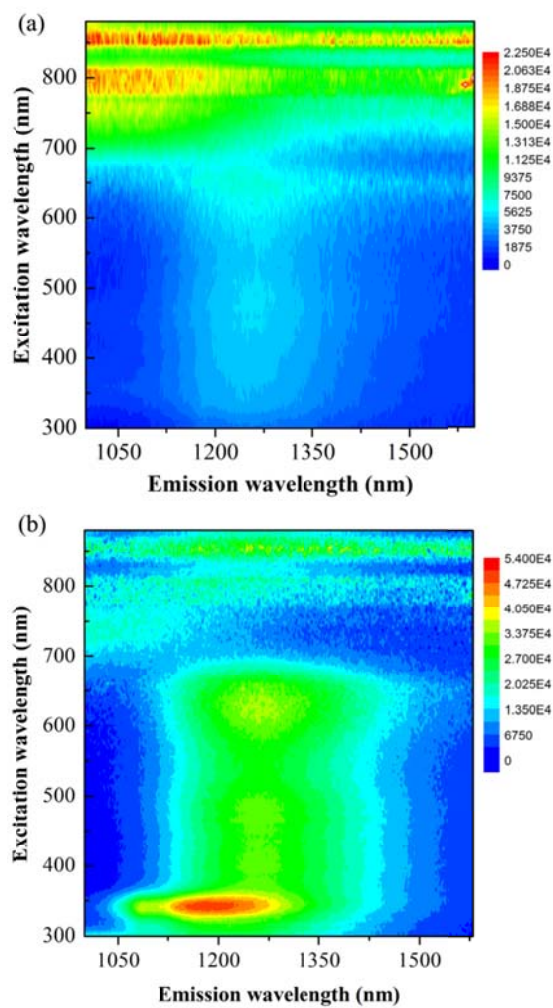

**Fig. 4** EEM spectra of (a) $Te_4(Ga_2Cl_7)_2$ and (b) $Te_4[Al_2Cl_7]_2$ products recorded at room temperature.

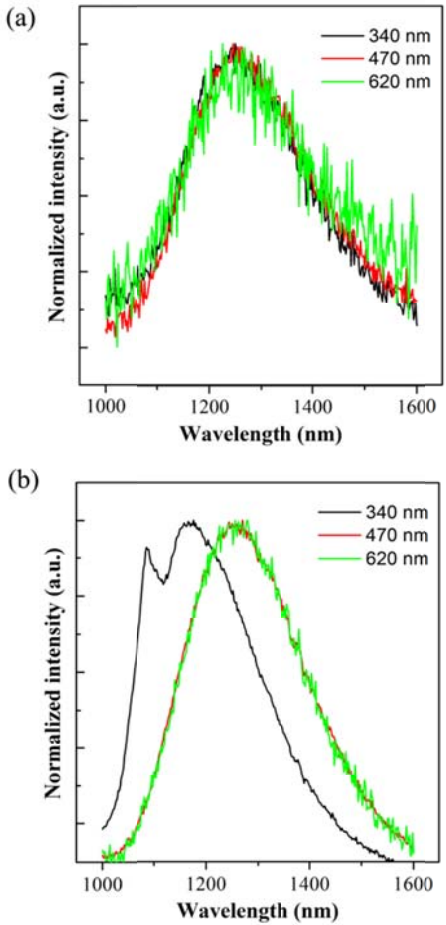

**Fig. 5** PL spectra of (a) Te$_4$(Ga$_2$Cl$_7$)$_2$ and (b) Te$_4$[Al$_2$Cl$_7$]$_2$ products under the excitation of 340, 470 and 620 nm.

In order to correlate the observed emission bands except those from Te$_4^{2+}$ with some Te species, next we carried out typical DFT calculations for the electronic structures and optical absorption spectra of some Te species including Te$_4$, Te$_4^+$, and Te$_4^{3+}$. Note the experimentally determined bond length of Te-Te and bond angle of Te-Te-Te in Te$_4$(Al$_2$Cl$_7$)$_2$ single crystals were used for the following calculations. All absorption spectra were fitted with a Gaussian function with a width at half-maximum of 50 nm. The ADF numerical integration parameter was set to 5.0 in all calculations. The calculated absorption spectra of Te$_4$, Te$_4^+$, and Te$_4^{3+}$ are displayed in Fig. 6. The theoretical excitation energies and oscillator strengths of these species are compiled in Tables S1−S3 (Supporting information). Both Te$_4$ and Te$_4^+$ display intense absorption in the wavelengh < 800 nm and relatively weak absorption in the NIR region. Te$_4$ demonstrates four NIR excitation bands at 2299, 2195, 1060, and 1041 nm, while Te$_4^+$ has two NIR bands located at 2349 and 2272 nm. It is interesting to note that Te$_4^{3+}$ has comparable oscillator strengths in the visible and NIR spectral regions and one forbidden-like mid-infrared electronic transition peaking at 7771 nm was also theoretically predicted. Based on the lines of evidences described above, it is clear that Te$_4$ and Te$_4^{3+}$ species have chances to emit in the range of 1000-1600 nm under the excitation of UV and visible lights (Fig. 6, and Table S1-S3). Although at present we cannot make exact assignments of the experimentally observed emission bands to these species, the aformentioned theoretical calculations suggest that it is reasonable to assign such emissions to netural or charged Te-related species. In addition to the visible and NIR photophysical behaviors of Te species, theoretical results also suggest that Te$_4$, Te$_4^+$, and Te$_4^{3+}$ units may emit at longer wavelengths since there are some electronic transitions over 2000 nm. Establishing the structure-property relationship of these species and developing new material systems containing such peculiar emitters will be the focus of our future studies.

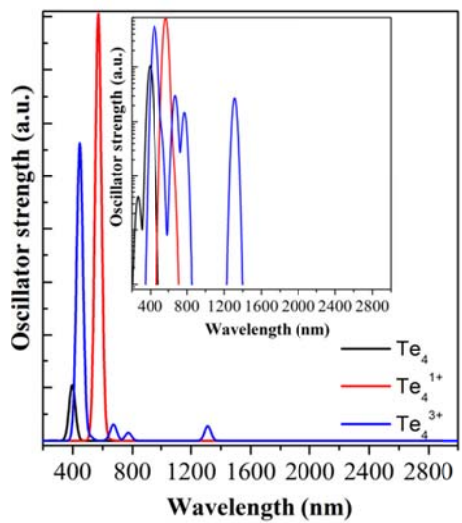

**Fig. 6** Theoretically calculated absorption spectra of $Te_4$, $Te_4^+$, and $Te_4^{3+}$. Inset: a logarithmic-scale version of the graph.

The combined evidences of XRD, absorption spectra, and steady-state PL results unambiguously make it clear that $Te_4^{2+}$ polycations emit in the NIR region with a peak at ca. 1260 nm. As is known, for a system containing more than one active center, the time-resolved PL display complex exponential decays because of the energy transfer beween adjacent centers. To know the PL dynamics of the obtained samples, we further took the decay curves by monitoring the signal at 1260 nm. As shown in Fig. 7, the curves of both samples greatly deviate from single-exponential decay, indicating that complicated energy migration processes occur in these systems. For $Te_4(Ga_2Cl_7)_2$ sample, a fast component, followed by a slow one, appears. In combination with the results discussed above, the former could be assigned to the fast energy transfer from $Te_4^{2+}$ to other optically active Te centers, while the latter may be attributable to the inherent electronic transition of $Te_4^{2+}$ from the excited level to the ground level. However, $Te_4(Al_2Cl_7)_2$ sample demonstrates different decay dynamics, possibly owing to the co-existence of $Te_4^{2+}$ and more other Te centers, as evidenced by XRD and steady-state PL results. Note that the experimental lifetimes are at least three orders of magnitude shorter than the theoretical values, for which the main reason might be the concentration quenching of $Te_4^{2+}$ species owing to the rather short spatical distance (< 1 nm) between them.

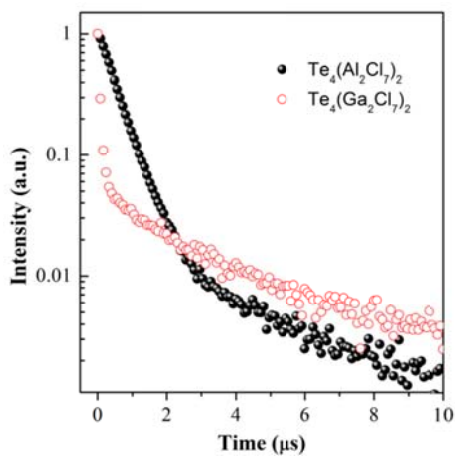

**Fig. 7** Decay curves monitored at 1260 nm for $Te_4(Ga_2Cl_7)_2$ and $Te_4[Al_2Cl_7]_2$ products under the excitation of 480 nm.

**Conclusion**

In conclusion, we have shown that molecular crystals containing $Te_4^{2+}$ polycations display broad NIR emission. The detailed experimental and theoretical results help us to confirm that $Te_4^{2+}$ units are peculiar NIR emitter with emission peaks at ca. 1260 nm and FWHMs over 240 nm. Moreover, it is found that both samples show excitation-wavelength-dependent emission lineshapes, implying that other active centers in addition to $Te_4^{2+}$ units exist. The following DFT calculations suggest that, most possibly, these centers are some charged or even neutral Te-related species. More detailed work is greatly needed to further clarify the origin of these emissions. This work not only provides insight on the PL mechanism of Te containing materials, but also evidences that charged Te clusters (e.g., $Te_4^{2+}$) with sizes less than 1 nm could be exploited as optically active centers, which bridge the gap between the well-dispersed emitters (e.g., rare earth and transition metal ions as well as color centers) and nanosized luminescent quantum dots. We hope that this could serve as a new design concept for developing new types of photonic materials, i.e, exploiting heavier p-block element polyhedra as emitters. It is also expected that such sub-nanometer species could exist in other systems such as glasses, polymers, and bulk optical crystals, and the successful stabilization of these centers in widely used hosts will greatly extend their practical applications, given the fact that the emission could cover important telecommunication and/or biological optical windows.

## Acknowledgements

H. Sun gratefully acknowledges the funding support from Hokkaido University and NIMS, Japan. H. Sun greatly thanks the fruitful discussion with Prof. M. Ruck in Dresden University of Technology, Germany, on the synthesis of high-quality crystals.

## Notes and references

[a] *Division of Materials Science and Engineering, Faculty of Engineering, Hokkaido University, Kita 13, Nishi 8, Kita-ku, Sapporo 060-8628, Japan. Fax: +81-11-706-7881; E-mail: timothyhsun@gmail.com*
[b] *International Center for Young Scientists (ICYS), National Institute for Material Sciences (NIMS), 1-2-1 Sengen, Tsukuba-city, Ibaraki 305-0047, Japan.*

[c] *Advanced Ceramics Group, Materials Processing Unit, National Institute for Materials Science (NIMS), 1-2-1 Sengen, Tsukuba-city, Ibaraki 305-0047, Japan*

[d] *PRESTO, Japan Science and Technology Agency (JST), 4-1-8 Honcho Kawaguchi, Saitama 332-0012, Japan*
[e] *Department of Electrical and Electronic Engineering, Kobe University, Kobe 657-8501, Japan*